\documentclass[3p,times,procedia]{elsarticle}
\usepackage{nupha_ecrc}

\volume{00}

\firstpage{1}

\journalname{Nuclear Physics A}

\runauth{}

\jid{nupha}

\jnltitlelogo{Nuclear Physics A}

\usepackage{amssymb}
\usepackage[figuresright]{rotating}

\begin{document}

\begin{frontmatter}

%% Title, authors and addresses

%% use the tnoteref command within \title for footnotes;
%% use the tnotetext command for the associated footnote;
%% use the fnref command within \author or \address for footnotes;
%% use the fntext command for the associated footnote;
%% use the corref command within \author for corresponding author footnotes;
%% use the cortext command for the associated footnote;
%% use the ead command for the email address,
%% and the form \ead[url] for the home page:
%%
%% \title{Title\tnoteref{label1}}
%% \tnotetext[label1]{}
%% \author{Name\corref{cor1}\fnref{label2}}
%% \ead{email address}
%% \ead[url]{home page}
%% \fntext[label2]{}
%% \cortext[cor1]{}
%% \address{Address\fnref{label3}}
%% \fntext[label3]{}

\dochead{}
%% Use \dochead if there is an article header, e.g. \dochead{Short communication}
%% \dochead can also be used to include a conference title, if directed by the editors
%% e.g. \dochead{17th International Conference on Dynamical Processes in Excited States of Solids}

\title{New insights from 3D simulations of heavy ion collisions}

%% use optional labels to link authors explicitly to addresses:
%% \author[label1,label2]{<author name>}
%% \address[label1]{<address>}
%% \address[label2]{<address>}

\author[l1]{Gabriel Denicol}

\author[l2]{Akihiko Monnai}

\author[l3]{Sangwook Ryu}

\author[l1]{Bj\"orn Schenke}

\address[l1]{Physics Department, Brookhaven National Laboratory, Upton, NY 11973, USA}
\address[l2]{RIKEN BNL Research Center, Brookhaven National Laboratory, Upton, NY 11973, USA}
\address[l3]{Department of Physics, McGill University, 3600 rue University, Montreal, Quebec H3A 2T8, Canada}

\begin{abstract}
Viscous relativistic hydrodynamics in 3+1 dimensions is applied to describe heavy ion collisions at RHIC and LHC. We present calculations of observables that are sensitive to the longitudinal structure of the created system.
In particular we present pseudo-rapidity correlations and demonstrate their dependence on both the initial state and short range correlations introduced via a microscopic transport description. We further demonstrate the effect of a varying temperature dependence of the shear viscosity to entropy density ratio on rapidity dependent flow harmonics.
\end{abstract}

%\begin{keyword}
%% keywords here, in the form: keyword \sep keyword

%% MSC codes here, in the form: \MSC code \sep code
%% or \MSC[2008] code \sep code (2000 is the default)

%\end{keyword}

\end{frontmatter}

%%
%% Start line numbering here if you want
%%
% \linenumbers

%% main text
\section{Introduction}
\label{sec:intro}
With the development of event-by-event 3+1 dimensional viscous relativistic hydrodynamic simulations \cite{Schenke:2010rr,Bozek:2011ua,Molnar:2014zha,Karpenko:2015xea}, a wide range of experimental observables in heavy ion collisions is now theoretically tractable. In particular multi-particle correlation measurements can be described using hydrodynamic calculations with certain initial conditions and allow for constraining important fundamental properties of quantum chromodynamic (QCD) matter at high temperatures.

Here we present calculations of two observables that require fully 3+1 dimensional simulations including fluctuations in the longitudinal direction. We employ the simulation \textsc{Music} \cite{Schenke:2010nt,Schenke:2010rr,Schenke:2011bn,Gale:2012rq} and a Monte Carlo (MC) Glauber type initial state that uses the Lexus model \cite{Jeon:1997bp,Monnai:2015sca} to determine fluctuating distributions of net baryon (quark) and entropy density distributions in rapidity.
For calculations of rapidity correlations we further employ a microscopic transport model (UrQMD) \cite{Bass:1998ca,Bleicher:1999xi} to take into account short range correlations introduced mainly by the kinematics of resonance decays.

\section{Model}
\label{sec:model}
The initial condition model for \textsc{Music} that we use here is described in \cite{Monnai:2015sca}. It consists of an MC Glauber model for constituent quarks, with their longitudinal positions determined from their rapidities. These rapidities follow from the initial $x$ value of the quark, sampled from a parton distribution function, and the change in rapidity for every quark-quark collision, determined from the Lexus model.
After smearing, the quark positions in space-time rapidity and transverse space directly determine the net baryon distribution. The entropy density distribution follows by placing tubes (stretching in rapidity) of constant entropy density between every quark and its last collision partner. This method produces a scaling of the multiplicity as in the number of quark participant model \cite{Eremin:2003qn}, which leads to a good description of the transverse energy distributions in both Au+Au and d+Au collisions \cite{Adler:2013aqf}.

The equation of state is constructed by interpolating the pressures of hadronic resonance gas and lattice QCD \cite{Borsanyi:2013bia,Borsanyi:2011sw} at 
a baryon chemical potential dependent connecting temperature \cite{Monnai:2015sca}. For temperatures around this value, the system crosses over into the quark gluon plasma (QGP) phase.
We evolve the system with \textsc{Music} using various scenarios for the temperature dependent shear and bulk viscosity to entropy density ratios, described below. Unless otherwise noted, we perform a Cooper-Frye freeze out at a constant energy density of $0.1\,{\rm GeV}/{\rm fm}^3$, compute thermal spectra of all particles with masses up to $2\,{\rm GeV}$, and decay resonances. Using the resulting particle spectra we calculate pseudo-rapidity correlations and rapidity dependent flow harmonics.

\section{Pseudorapidity correlations}
The ATLAS collaboration has presented results on two-particle pseudorapidity correlations, expanded into Legendre polynomials \cite{ATLAS-CONF-2015-020,ATLAS-CONF-2015-051}. The Legendre coefficients contain information on the event-by-event particle production mechanism, in particular the so far largely unknown structure in pseudorapidity.
In \cite{Monnai:2015sca} we used the model described above to study the effect of 1) the number of sources and 2) the transport properties of the medium on the Legendre coefficients
\begin{equation}
  a_{n,m} = \int_{-Y}^{Y} C_N(\eta_1,\eta_2) \frac{T_n(\eta_1) T_m(\eta_2) + T_n(\eta_2) T_m(\eta_1)}{2} \frac{d\eta_1}{Y} \frac{d\eta_2}{Y}\,,
\end{equation}
where $T_n(\eta_p) = \sqrt{n+1/2}\,P_n(\eta_p/Y)$, $P_n$ are the standard Legendre polynomials, and $C_N$ is the two-particle correlation function with  the effect of residual centrality dependence in the average shape $\langle N(\eta) \rangle$ removed \cite{Jia:2015jga}.

\begin{figure}[htb]
   \begin{center}
     \includegraphics[width=7.4cm]{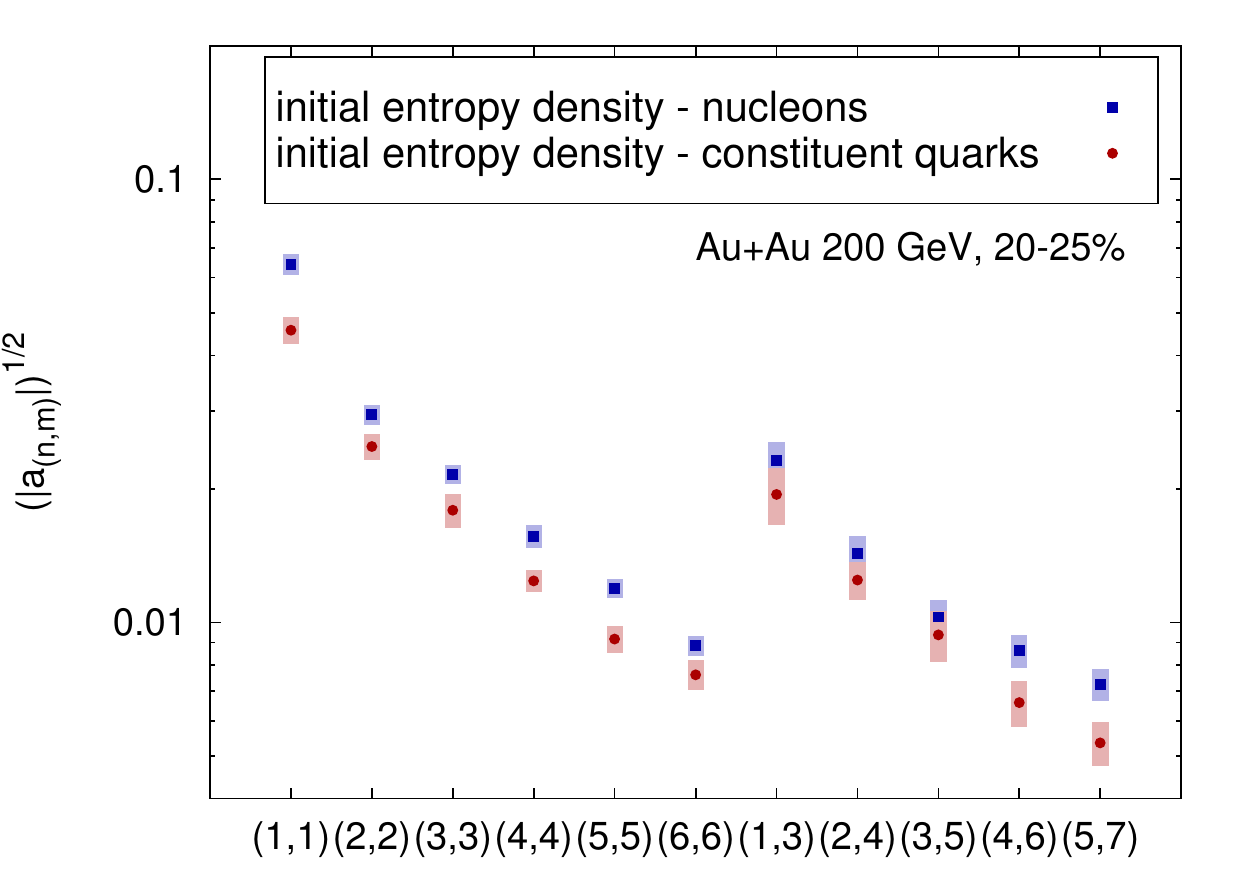}
     \includegraphics[width=7.4cm]{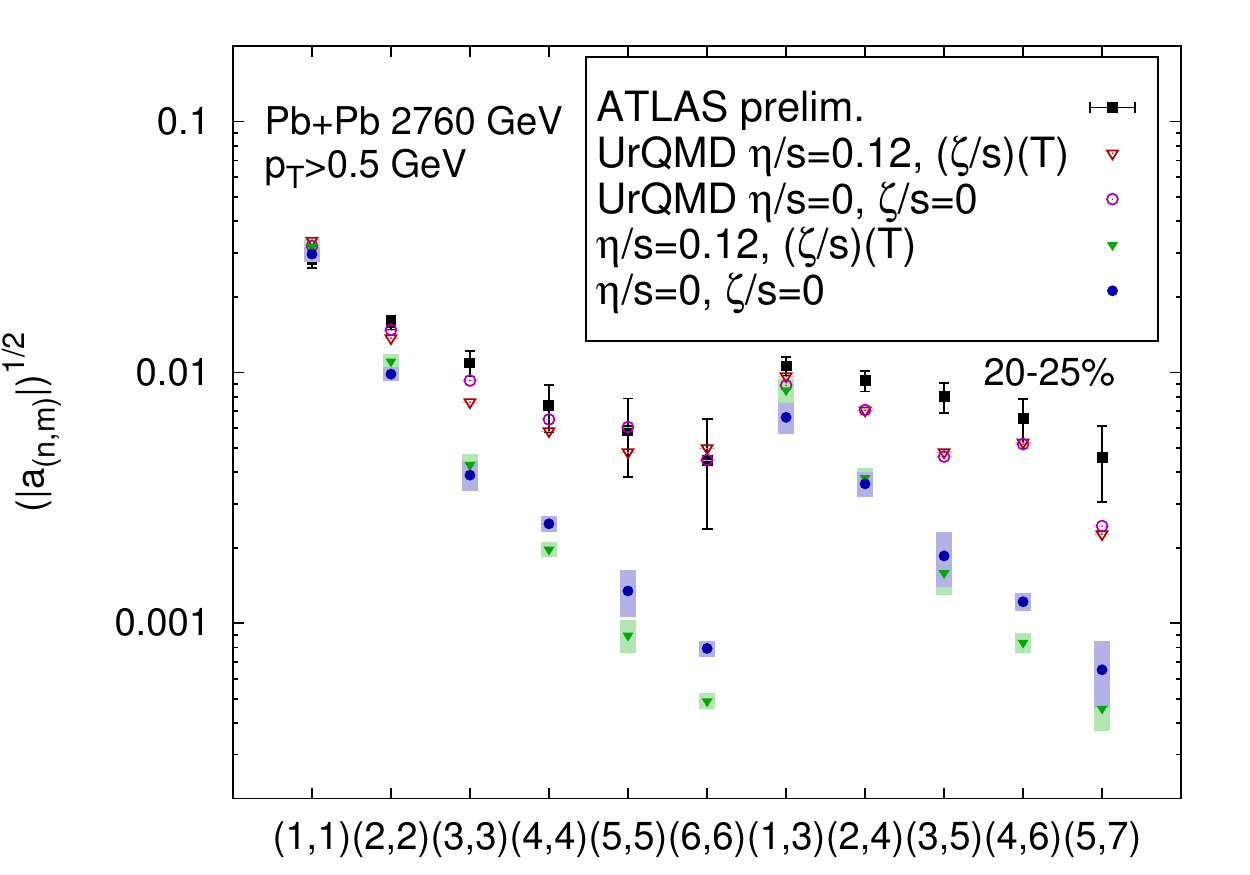}
     \caption{Left: Legendre coefficients $\sqrt{|a_{n,m}|}$ from the initial entropy density distribution. Figure from \cite{Monnai:2015sca}.
       Right: Legendre coefficients $\sqrt{|a_{n,m}|}$ after hydrodynamic evolution using average spectra from Cooper Frye for ideal hydro (filled circles) and viscous hydro (filled triangles). Open symbols show the same calculations but with statistical hadronization and UrQMD evolution, which keeps the short range correlations. Experimental data from the ATLAS collaboration \cite{ATLAS-CONF-2015-020}.
}
     \label{fig:anm-init}
   \end{center}
   \vspace{-0.5cm}
\end{figure}

The number of particle producing sources should have a measurable effect as shown for the $a_{n,m}$ extracted from the initial state entropy density distributions in Fig.\,\ref{fig:anm-init} (left), where we compare the constituent quark model described above to the same model, but with nucleon degrees of freedom. In the latter, we have fewer sources, leading to more fluctuations and thus larger $a_{n,m}$.

It was shown in \cite{Monnai:2015sca} that the effect of transport parameters on the $a_{n,m}$ is fairly weak, compared to potential short range non-flow correlations that were not included (also see Fig.\,\ref{fig:anm-init} (right)). Here we explicitly include such non-flow effects by performing a statistical hadronization, similar to \cite{Bozek:2015tca}. We use the microscopic hadronic transport model UrQMD \cite{Bass:1998ca,Bleicher:1999xi}, to describe the low temperature evolution of the system and compute the $a_{n,m}$ from the final hadron distributions. The result is shown in Fig.\,\ref{fig:anm-init} (right), where one can see that the short range non-flow correlations are the dominant effect for all $a_{n,m}$ with $n,m>1$. Their inclusion leads to much better agreement with the experimental data \cite{ATLAS-CONF-2015-020}.
We note that in the meantime the ATLAS collaboration has introduced a method to remove short range correlations \cite{ATLAS-CONF-2015-051}. Doing so, all coefficients except $a_{1,1}$ are consistent with zero within experimental errors, which seems consistent with the pure hydrodynamic results, where all higher $a_{n,m}$ are largely suppressed.

\section{Rapidity dependent flow harmonics and the shear viscosity of QCD matter}
\begin{figure}[htb]
\includegraphics[width=0.5\textwidth]{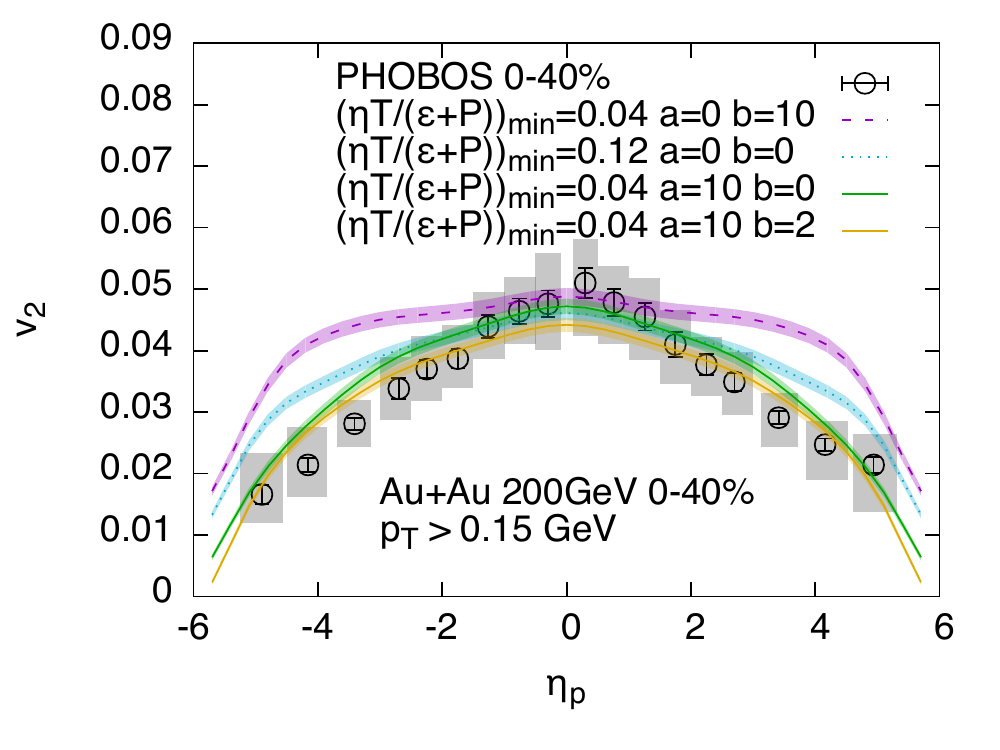}
\includegraphics[width=0.5\textwidth]{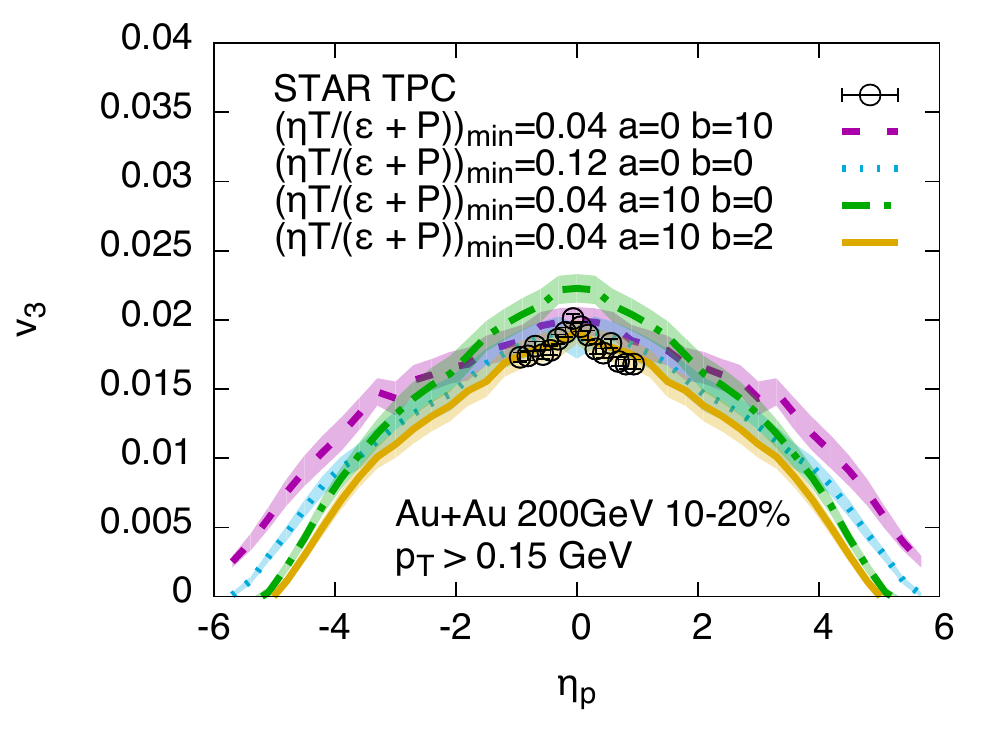}
\caption{Left: $v_2$ of charged hadrons at 0-40\% centrality as a function of pseudo-rapidity for four different shear viscosity scenarios compared to experimental data from the PHOBOS collaboration \cite{Back:2004zg}. Figure from \cite{Denicol:2015nhu}. Right: Same as the left figure but for $v_3$ in $10-20\%$ centrality, compared to experimental data from the STAR collaboration \cite{Adamczyk:2013waa}.\label{fig:v2-eta}}
\end{figure}

In the same framework as described in Section \ref{sec:model} we compute the charged hadron flow harmonics $v_n$ as functions of pseudorapidity.
This calculation was first presented in \cite{Denicol:2015nhu}. In this study we focus on the temperature dependence of $\eta/s$ (the bulk viscosity is also temperature dependent but held fixed as described in \cite{Denicol:2015nhu}).
We employ the following simple parametrization of $\eta T/(\varepsilon+P)$, which is close to the shear viscosity to entropy density ratio for small $\mu_B$:
\begin{equation}
(\eta T/(\varepsilon+P))(T) = (\eta T/(\varepsilon+P))_{\rm min} + a \times (T_c-T)\theta(T_c-T) + b \times (T-T_c)\theta(T-T_c)\,.
\end{equation}
As shown in Fig.\,\ref{fig:v2-eta}, different scenarios of temperature dependent $\eta/s$, can lead to significantly different $v_n$ at large rapidities.
We find that agreement with the data requires 1) a large hadronic viscosity 2) a small minimum, possibly lower than $1/4\pi$, and 3) a weak increase of $\eta/s$ in the QGP regime. 

Our calculations also show that the widths of the $v_n$ event-by-event distributions are largely independent of the transport parameters. In combination with precise measurements over a wide range in rapidity this fact can be used to constrain the three dimensional fluctuating initial state.

\section{Conclusions}
We have presented results for different observables that have become theoretically accessible with the development of 3+1 dimensional viscous relativistic hydrodynamics and initial state models that provide fluctuations in three spatial dimensions. Pseudorapidity correlations are sensitive to the number of sources and thus to the detailed particle production mechanism in heavy ion collisions. They are also very sensitive to short range correlations included in microscopic transport simulations. The study of flow harmonics and their distributions at forward rapidities has a strong potential to constrain the temperature dependent transport properties of QCD matter. We have presented first constraints obtained from comparison to existing RHIC data.

%% The Appendices part is started with the command \appendix;
%% appendix sections are then done as normal sections
%% \appendix

%% \section{}
%% \label{}

%% References
%%
%% Following citation commands can be used in the body text:
%% Usage of \cite is as follows:
%%   \cite{key}         ==>>  [#]
%%   \cite[chap. 2]{key} ==>> [#, chap. 2]
%%

\section*{Acknowledgments}
We thank Charles Gale and Sangyong Jeon for useful discussions. AM is supported by the RIKEN Special Postdoctoral Researcher program. GSD and BPS are supported under DOE Contract No. DE-SC0012704. S.R. is supported in part by the Natural Sciences and Engineering Research Council of Canada. This research used resources of the National Energy Research Scientific Computing Center, which is supported by the Office of Science of the U.S. Department of Energy under Contract No. DE-AC02-05CH11231, and the Guillimin supercomputer at McGill University under the auspices of Calcul Quebec and Compute Canada. The operation of Guillimin is funded by the Canada Foundation for Innovation (CFI), the National Science and Engineering Research Council (NSERC), NanoQuebec, and the Fonds Quebecois de Recherche sur la Nature et les Technologies (FQRNT). BPS acknowledges a DOE Office of Science Early Career Award.

%% References with BibTeX database:

\bibliographystyle{elsarticle-num}
\bibliography{spires}

%% Authors are advised to use a BibTeX database file for their reference list.
%% The provided style file elsarticle-num.bst formats references in the required Procedia style

%% For references without a BibTeX database:

% \begin{thebibliography}{00}

%% \bibitem must have the following form:
%%   \bibitem{key}...
%%

% \bibitem{}

% \end{thebibliography}

\end{document}